\documentclass[reprint, twocolumn, amsmath, amssymb, showpacs, superscriptaddress, aps, prl]{revtex4-2}

\usepackage{microtype}
\usepackage{graphicx}
\usepackage{dcolumn}
\usepackage{bm}
\usepackage[colorlinks=true,
						linkcolor=blue,
						urlcolor=blue,
						citecolor=blue,
						bookmarks=true,
						pdfborder={0 0 0}]{hyperref}
\usepackage{multirow}
\usepackage[normalem]{ulem}
\usepackage{xcolor}
\usepackage[export]{adjustbox}

\graphicspath{{figures/}}

\usepackage{url}

\begin{document}
\title{Model of spatial competition on discrete markets}
\author{Andrea Civilini}
\affiliation{School of Mathematical Sciences, Queen Mary University of London, London E1 4NS, United Kingdom}
\author{Vito Latora}
\affiliation{School of Mathematical Sciences, Queen Mary University of London, London E1 4NS, United Kingdom}
\affiliation{Dipartimento di Fisica ed Astronomia, Universit\`a di Catania and INFN, Catania I-95123, Italy}
\affiliation{Complexity Science Hub Vienna, A-1080 Vienna, Austria}

\begin{abstract}
We propose a dynamical model of price formation on a spatial market 
where sellers and buyers are placed on the nodes of a graph, and the 
distribution of the buyers depends on the positions and prices of the sellers. 
We find that, depending on the positions of the sellers and on the level of information available, the price dynamics of our model can either converge to fixed prices, or produce cycles of different amplitudes and periods. 
We show how to measure the strength of competition in a spatial network by
extracting the exponent of the scaling of the prices with the size of the system. As an application, we characterize the different level of
competition in street networks of real cities across the
globe. Finally, using the model dynamics we can define a novel measure of node centrality, which quantifies the relevance of a node in a competitive market.

\end{abstract}

\maketitle

Oligopoly pricing, i.e.~how market prices are formed in the presence of two or more competitors, is a central topic in economic theory. 
The origin of the interest in oligopoly pricing can be tracked back to a series of classical papers from the end of 19th century, which can be regarded as the ancestors of modern non-cooperative game theory~\cite{vives2001}.
In 1883, Bertrand showed that, in the presence of price competition, the Cournot model of duopoly~\cite{cournot1838}
led to the famous Bertrand's Paradox, unrealistically predicting perfect competition (i.e. zero profit) even in the
simple case of two sellers~\cite{bertrand1883}. 
A first solution to the Bertrand's paradox was proposed by Edgeworth in 1925: by introducing sellers with capacity constraint, i.e. such that each one of the sellers alone is unable to serve the whole market, he showed the emergence of the so-called Edgeworth price cycles, oscillations in the prices having positive profit for the sellers~\cite{edgeworth1925}. 
In his seminal 1929 work, Hotelling first noticed how the market structure can have a great impact on the stability of prices in an oligopoly~\cite{hotelling1929}. 
He introduced a model where two sellers occupy different positions on a spatial market represented as a line segment. 
Hotelling showed how in his model  prices converge to non-trivial fixed values with positive profit as a consequence of the market structure and of the positions occupied by the sellers. 
Even if qualitatively opposite conclusions can be drawn from the solutions proposed by Edgeworth and Hotelling to the Bertrand's paradox, both fixed (\`{a} la Hotelling) and cycling (\`{a} la Edgeworth) prices with positive profit have indeed been observed in the  real world.
In particular, thanks to the increasing amount of fine grained prices' data available, Edgeworth price cycles, once thought a purely theoretical construction~\cite{dudey1992}, have been recently observed in real-world markets such as the retail gasoline market and online bid platforms~\cite{noel2008edgeworth, zhang2011reverseedge}.
Despite these fundamental results, a full understanding of the link between price dynamics and the spatial structure of a market is still missing. Firstly, due to its mathematical complexity, a general solution of the Hotelling model on 
a line segment has not yet been found~\cite{vickrey1964, vickrey1964_1999, daspremont1979, GALOR1982, dasgupta1986, osborne1987, biscaia2013_review}.
Moreover, a line segment is a too strong and unrealistic assumption~\cite{vickrey1964, vickrey1964_1999}, and this greatly limits the  applicability of the Hotelling model to real-world spatially structured markets. 
Finally, it is also challenging to incorporate in the Hotelling model realistic aspects of the decision making process, such as bounded rationality and cost of information~\cite{simon1955, stigler1961}, which have been proved to play a central role in many real-world systems~\cite{Funk2009TheSO, david_boundedAI2015, bruch2015, ZHAN2018}.

In this Letter, we propose a dynamical model of competition in which sellers and buyers are placed on the nodes of a graph representing any arbitrarily complex market structure. The main ingredient of the model is the mutual influence between buyers' location and sellers' price dynamics.  
Namely, the spatial distribution of buyers is not fixed, but dynamically depends on the prices and positions of sellers. At the same time, the price dynamics is influenced by the distribution of the buyers.
In this way, our model allows to include different price update rules and also buyers with limited information about the market (bounded rationality), by naturally modelling them as random walkers.
We find that, depending on the positions of the sellers on the graph, and on the level of information available, 
the price dynamics of our model can either converge to fixed prices (\`{a} la Hotelling), or produce Edgeworth cycles of different amplitudes and periods. 
As an additional benefit, our model allows to study how maximum prices scale with the size of a market. By extracting the scaling exponent, we can then define the ``market competition dimension'' of a spatial system. As an application, we characterize the different level of competition in street networks of real cities across the globe, comparing them to regular two-dimensional square lattices with the same number of nodes. 

\begin{figure}[htp]
    \includegraphics[width=0.5\textwidth]{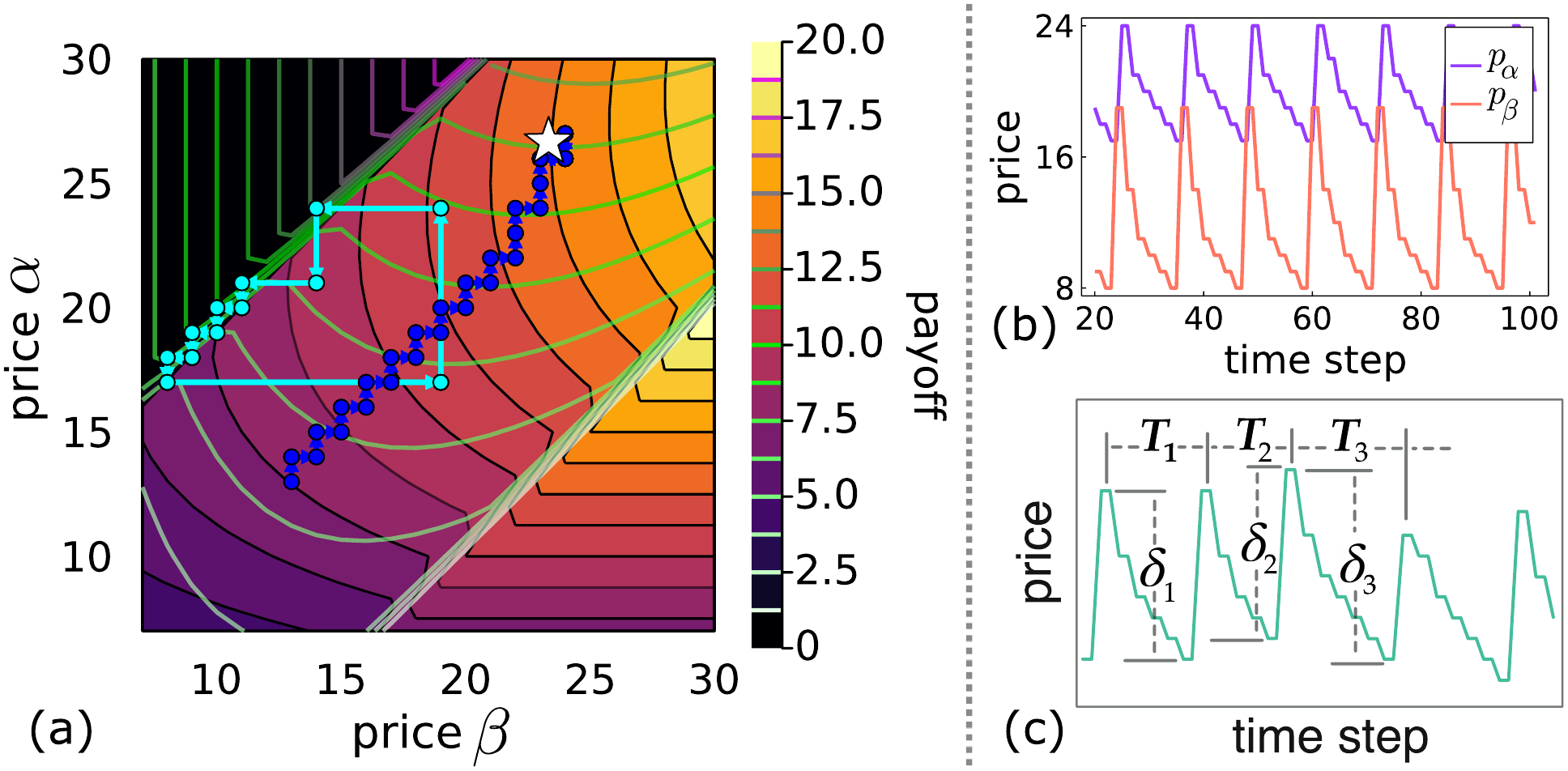}
  \caption{Price dynamics on a chain of $N = 25$ nodes, with seller positions $n_{\alpha} = 10$ and $n_{\beta} = 20$. 
  (a) The heatmap shows the payoff of seller $\alpha$, the payoff's level curves of seller $\beta$ are represented in green-magenta.
  The blue and cyan trajectories show the prices dynamics under a One-Step (OS) and Best Response (BR) update respectively. 
  The OS converges to the equilibrium prices predicted by the continuous Hotelling linear model (white star), while the cyan trajectory shows price cycles. (b) Price cycles over time.
  (c) For each cycle $i$ we measured the period $T_i$, the max and min prices and the amplitude $\delta_i$.
  }
\label{fig:1d_market}
\end{figure}

\paragraph*{The model.}

In our model the market units are the nodes of an undirected graph~\cite{newmannetwork}, whose links can represent either geographical adjacency or feature similarity. Sellers and buyers are placed on the nodes of the graph. 
We focus here on the case of two sellers $\alpha$ and $\beta$, selling a commodity respectively at prices $p_{\alpha}$ and $p_{\beta}$, 
although the model can be readily extended to more sellers. The process by which buyers 
search for the seller to buy from is modelled as a random walk, that combines two different mechanisms. With a probability $w$, a buyer at node $i$ is fully informed about positions and prices of the sellers.
Hence, at each time step, it moves on the graph towards the seller $\alpha$ with the lower delivered price ${\cal P}_{i \alpha}$:
\begin{equation} \label{eq:delivered_price}
 {\cal P}_{i \alpha} = d_{i \alpha} + p_{\alpha}
\end{equation}
where $d_{i \alpha}$ is the distance from node $i$ to seller $\alpha$.
Instead, with probability $1-w$, a buyer has bounded rationality and moves as a lazy random walker, i.e. it either jumps on one of the neighbours with uniform probability or it remains on the same node, 
as long as the price difference $|p_{\alpha} - p_{\beta}|$ is below a given threshold $\Delta p_{T}$ (representing the limit of bounded rationality). 
The row vector $\bm{\phi}(t)= \{ \phi_i (t) \}_{i=1,\ldots,N}$, representing the distribution of buyers, evolves over time according to $\bm{\phi}(t) = \bm{\phi}(t-1) \Pi $, where the elements of the transition matrix $\Pi$ are:
\begin{align}
    \Pi_{ij} &= w\left[ a_{i j} \frac{ \sum_{\alpha} \sigma_{i \alpha}(j) \Theta(\min_{\beta}({\cal P}_{i \beta}) - {\cal P}_{i \alpha})}{ \sum_{\alpha} \sigma_{i \alpha} \Theta(\min_{\beta}({\cal P}_{i \beta}) - {\cal P}_{i \alpha})}\right] \nonumber
    \\
     &+ (1 - w )\left( \frac{a_{i j}}{2}\frac{1}{k_i} + \frac{\delta_{i j}}{2} \right)
\end{align}

Here, $A=\{ a_{i j} \}$ is the adjacency matrix of the graph,
$\sigma_{i \alpha}(j)$ is the number of shortest paths from node $i$ to the seller $\alpha$ passing through node $j$, while $\sigma_{i \alpha}$ is the total number of shortest paths. 
The function $\Theta$ is the Heaviside step function, such that $\Theta(x) = 1,$ for $x \geq 0$ otherwise $\Theta(x) = 0$, while $\delta_{ij}$ is the Kronecker delta. 
Since the Markov chain is aperiodic and the graph is connected, a stationary distribution $\bm{\phi}^*= \bm{\phi}^* \Pi$ always exists unique~\cite{levin2017markov} and can be easily found using a standard power method~\cite{ANDRILLI2016}.
Once the buyers have reached their stationary distribution, the two sellers evaluate their payoffs $\pi_{\alpha} = \phi^*_{\alpha}p_{\alpha}$  and $\pi_{\beta} = \phi^*_{\beta}p_{\beta}$.  
A randomly selected seller, let's say $\alpha$,
can then change its price $p_{\alpha}$ in order to increase its payoff.  
We considered two different price update rules: in the
One-Step (OS) dynamics the seller slowly adjusts its price in steps of $\Delta_p = \pm 1$, while in the Best Response (BR) it can 
choose the new $p_{\alpha}$ theoretically among all possible prices  (practically, we restricted the price range to two times the graph diameter).
A new stationary distribution of buyers is then recalculated according to the new value of $p_{\alpha}$, i.e. the new delivered prices in Eq.~\eqref{eq:delivered_price}.
If the payoff associated with the new $p_{\alpha}$
is higher than the old payoff, the seller will adopt the new price, otherwise the seller will keep the old one.
Hence, the price dynamics is driven by two opposite forces: while on the one hand increasing (decreasing) the price $p_{\alpha}$
will lead seller $\alpha$ to earn more (less) from each buyer, on the other hand it will reduce (increase) the number of sellers buying from seller $\alpha$.

\begin{figure}[htp]
    \includegraphics[width=0.5\textwidth]{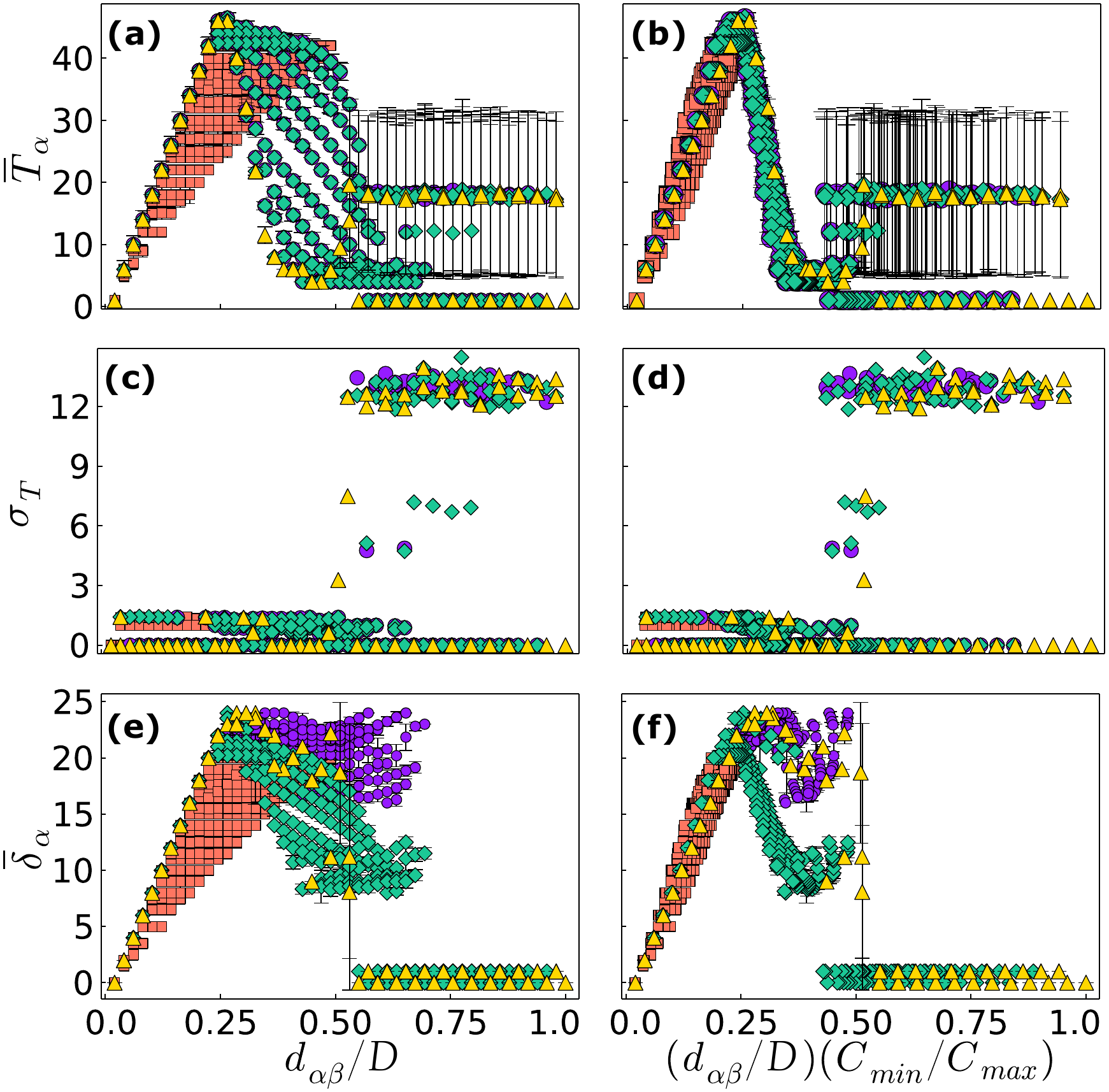}
  \caption{Average period and variance $\overline{T}$ and $\sigma_T$, and average amplitude $\overline{\delta}$ of the BR dynamics on a chain of $N = 50$ nodes as a function of the normalized seller distances $d'_{\alpha\beta}$ (left panels) and $d''_{\alpha\beta}$ (right panels). 
 Denoting the chain's middle point $x_c = 25.5$ and the sellers centre of mass $n_c = (n_{\alpha} + n_{\beta})/2$, the orange squares are for positions $(n_{\alpha}, n_{\beta})$ such that $n_{\alpha}, n_{\beta} < x_c$, the purple circles, green diamonds and yellow triangles are for $n_{\alpha} < x_c < n_{\beta}$ with $n_c < x_c - 0.5$, $n_c > x_c + 0.5$ and $|n_c - x_c| \leq 0.5$ respectively.
  The vertical bars are the standard deviations. 
  }
\label{fig:edgeworth_period_sellers_positions}
\end{figure}

\begin{figure}[htp]
    \includegraphics[width=0.5\textwidth]{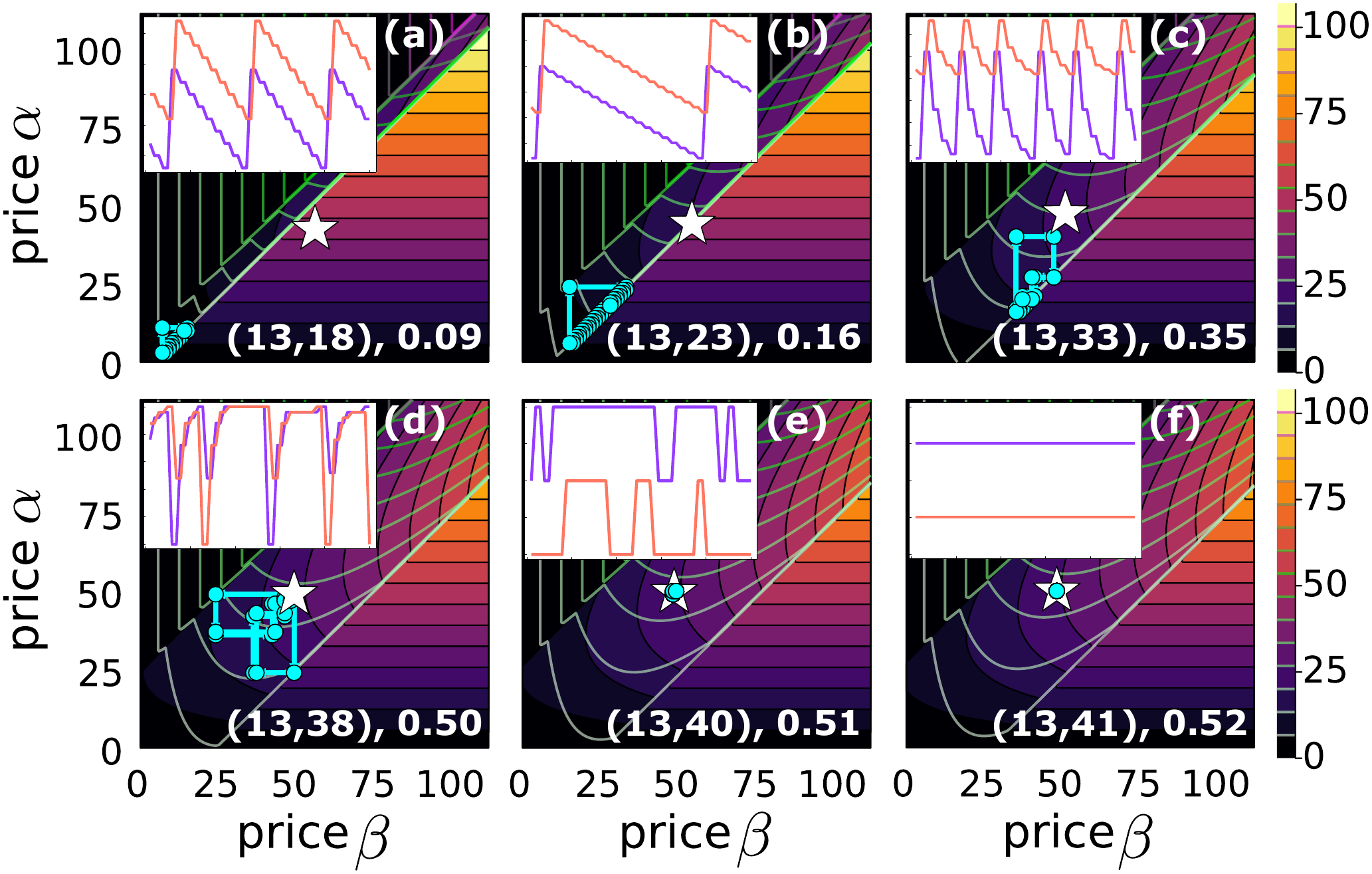}
  \caption{Price trajectories for increasing sellers distance, with the values $(n_{\alpha}, n_{\beta})$, $d''$ reported in each panel.  
  For normalized distance $d''$ lower than $\sim 0.5$ we 
  we observe Edgeworth's cycles with (a), (b) $\overline{T}$ and $\overline{\delta}$ first increasing with the distance (for 
  $d''<  0.25$) and then (c) decreasing 
  with it  (for $0.25 < d'' < 0.5$). 
    (d) Irregular reverse cycles for $d'' \approx 0.5$. (e), (f) For $d''>0.5$ we can either have Hotelling fixed points or irregular oscillations around the fixed points.}
\label{fig:edgeworth_cycles_zoology}
\end{figure}

\paragraph*{Results.}
We first characterize the price dynamics in the simplest case in which the sellers are placed on a chain of $N$ nodes and the buyers have perfect information, i.e. $w=1$.  
When the sellers occupy two nodes $n_{\alpha}$ and $n_{\beta}$ at distance $d_{\alpha \beta}$, we expect to recover the same 
equilibrium prices, corresponding to a Nash Equilibrium~\cite{acourseinGT} of the original Hotelling model~\cite{hotelling1929}:
\begin{align}
    p^H_{\alpha} = N + \frac{a - b}{3},
    \qquad 
    p^H_{\beta} = N - \frac{a - b}{3} \label{eq:pa}
\end{align}
when $N$ and $d_{\alpha \beta}$ are sufficiently larger than $1$, i.e. in the continuous limit. Here $a$ and $b$ are the distances of the two sellers from the closer extremity of the chain.  
Fig.~\ref{fig:1d_market} shows the price dynamics of our model when the two sellers are at distance $d_{\alpha \beta} =10$ on a chain of $N=25$ nodes. Under the One-Step (OS) price update rule, the dynamics converges to a local maximum of the two seller payoffs, 
which corresponds to the Hotelling fixed price solutions in Eqs.~\eqref{eq:pa}. We have verified that fixed prices are recovered for all the positions of the two sellers, except when they are too close and both on the same side of the chain (see SM ). 
If instead the sellers update the prices according to a Best Response (BR), i.e. without any restriction on the price step  $\Delta p$, our model is able to produce 
Edgeworth price cycles~\cite{edgeworth1925, maskin1988}.
In this case, the prices form closed stationary cycles with a peculiar asymmetric pattern: the two sellers start undercutting the prices until a lower price bound, when one of the seller increases the price up to an upper price threshold, immediately followed by the other seller, and then the price undercutting process starts again. 
In Fig.~\ref{fig:edgeworth_period_sellers_positions} 
we characterize the features of the BR price dynamics as a function of the positions of the two sellers on the chain.
In particular, we report the average cycle period $\overline{T}$ and its variance $\sigma_T$, and the average amplitude $\overline{\delta}$ 
as functions of the normalized distances  
$d'_{\alpha\beta} = d_{\alpha \beta} /D$, here $D$ is
the chain diameter, and $d''_{\alpha\beta} = d'_{\alpha\beta} C_{min} / C_{max}$, 
where $C_{min}$ and $C_{max}$ are the smallest and largest values between 
the closeness centralities $C_{n_{\alpha}}$ and $C_{n_{\beta}}$  of the two sellers. 
According to the sellers positions, the model gives rise to four different types of stationary solutions, whose dynamics is illustrated in Fig.~\ref{fig:edgeworth_cycles_zoology}. 
  For values of $d'' < 0.5$ we observe Edgeworth's cycles (Fig.~\ref{fig:edgeworth_cycles_zoology}a) with average cycle period $\overline{T}$ and average amplitude $\overline{\delta}$ (Fig.~\ref{fig:edgeworth_period_sellers_positions}b,f) that increase with $d''$ until $d'' \approx 0.25$ (Fig.~\ref{fig:edgeworth_cycles_zoology}b), and then decrease until $d'' \approx 0.5$ (Fig.~\ref{fig:edgeworth_cycles_zoology}c) . 
  For $d'' > 0.5$, we have the co-existence of Hotelling fixed points (Fig.~\ref{fig:edgeworth_cycles_zoology}f) and irregularly oscillating solutions (with fluctuations of the order of the price unit) around Hotelling fixed points (Fig.~\ref{fig:edgeworth_cycles_zoology}e). 
  The appearance of irregular oscillations at $d'' \approx 0.5$ is revealed by the sharp change in 
   the variance of the periods $\sigma_T$ in Fig.~\ref{fig:edgeworth_period_sellers_positions}c,d.
  Finally, for $d'' \approx 0.5^-$, together with the two solutions found for $d'' > 0.5$, we also observe the emergence of price cycles with broad amplitude as in Edgeworth cycles but irregular/chaotic period as for the solutions oscillating around the Hotelling equilibrium (Fig.~\ref{fig:edgeworth_cycles_zoology}d). 
  In such cycles the two sellers follow a pattern that is inverted to that of the standard Edgeworth cycles.   
  They spend most of the time at the Hotelling equilibrium, until one of the two sellers dramatically cuts its price to steal the buyers from the other seller, still making the same profit as before. The other seller sees its earning dropping to zero and is then forced to follow the other seller in cutting the price to reacquire part of the market. Then, the two sellers start rising the prices, until they reach again the Hotelling equilibrium. Notice that {\em reverse Edgeworth cycles}, such as those theoretically predicted by our model, can indeed exist in real systems, as recently found in empirical analyses of online bid platforms~\cite{zhang2011reverseedge}. 
Let us now consider the case $w<1$ where a fraction of the buyers has no information on seller positions and prices. For the sake of simplicity and with no lack of generality, we will present here the symmetric case $n_{\alpha} = n_{\beta}$, which allows a complete analytical characterization of the price dynamics (for $w<1$ and $n_{\alpha} \neq n_{\beta} $ see SM). 
Edgeworth cycles also emerge for $w<1$, 
depending on the values of $w$ and of the limit of bounded rationality $\Delta p_{T}$.  In particular, we observe a critical value of $w_c= 1 - 2/(2 + \Delta p_{T})$, such that Edgeworth cycles exist only for $w < w_c$, while the minimum and maximum prices in each cycle are given by (see SM for details):
\begin{align}
    p_{m} = \frac{(1-w)\Delta p_{T} + 1}{2w} + \frac{1}{2},
    \quad 
    p_{M} = p_{m} + \Delta p_{T}
\end{align}

\begin{figure}[htp]
    \includegraphics[width=0.5\textwidth]{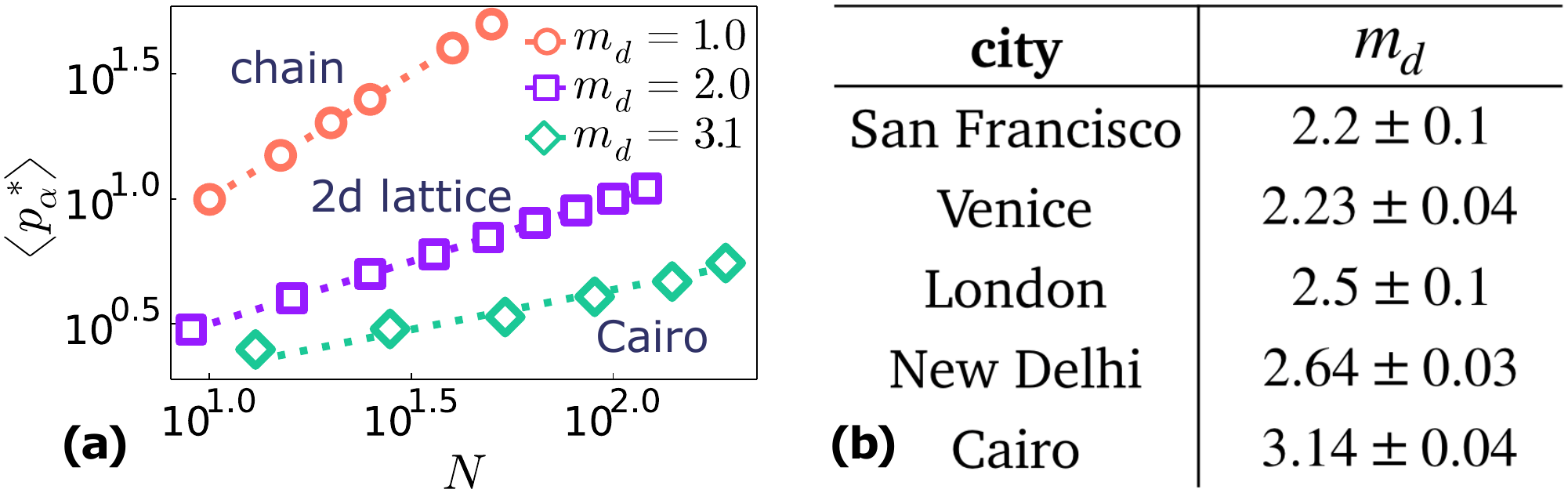}
  \caption{(a) Maximum values $\langle p_{\alpha}^* \rangle$ of the average equilibrium price as function of the graph order $N$ for different networks: a chain, a regular square lattice and the real street network of a city (in this plot Cairo, Egypt). We report the corresponding market competition dimension $m_d$ obtained by fitting $\langle p_{\alpha}^* \rangle(N) = N^{1/m_d}$ to the data points. (b) $m_d$ values for several cities. The reported error is the standard error of the estimate.}
\label{fig:market_dimension}
\end{figure}

Finally, in order to explore the effects that the topology
of the market has on the price dynamics, we have studied
our model on more complex graphs, in the case $w=1$ of a perfectly rational population of buyers. 
As examples of real-world markets we have considered the spatial structures of urban street networks of $19$ 
cities from all over the world~\cite{crucitti2006,cardillo2006}. 
We focus on the average equilibrium price, $\langle p_{\alpha}^*\rangle_{d_{\alpha \beta}}$, where $\langle \cdot  \rangle_{d_{\alpha \beta}}$  indicates the average over all the positions of the two sellers for a fixed distance $d_{\alpha \beta}$. 
Our numerical simulations show that, for sufficiently large values of  $d_{\alpha \beta}$, the 
average equilibrium price reaches a maximum value ${\langle p_{\alpha}^*\rangle}$ (see SM), which is a function of the graph size $N$ and of the topology of the network, but does not depend on the type of price dynamics (namely OS or BR). 
Fig.~\ref{fig:market_dimension}a shows the scaling of ${\langle p_{\alpha}^*\rangle}$ with $N$ for different networks. In particular, we compare a linear market (chain) to a 2-dimensional square lattice and to the street network of Cairo. 
By fitting the numerical results by $\langle p_{\alpha}^* \rangle(N) \sim N^{1/m_d}$, it is possible to define the \emph{competition dimension} of a market from the value of $m_d$. %
In fact, the exponent $m_d$ quantifies how the maximum price (and payoff) increases with the order of the graph $N$. The higher $m_d$, the slower the maximum price increases, i.e. $m_d$ is a measure of the price competition.
In particular, we obtained a value of $m_d = 1$ for the linear market and $m_d = 2$ for a square lattice.
Hence, it is intuitive to identify $m_d$ as the dimension of a spatial market. 
Fig.~\ref{fig:market_dimension}b reports the values of $m_d$ obtained for the street networks of different cities (see SM for more details and for other cities).
Interestingly, for all real-world street networks we found $2 < m_d < 3.2 $ 
(see SM), i.e. the price competition is enhanced with respect to a 2-d square lattice, and the maximum prices increase considerably slower. 

\begin{figure}[htp]
    \includegraphics[width=0.5\textwidth]{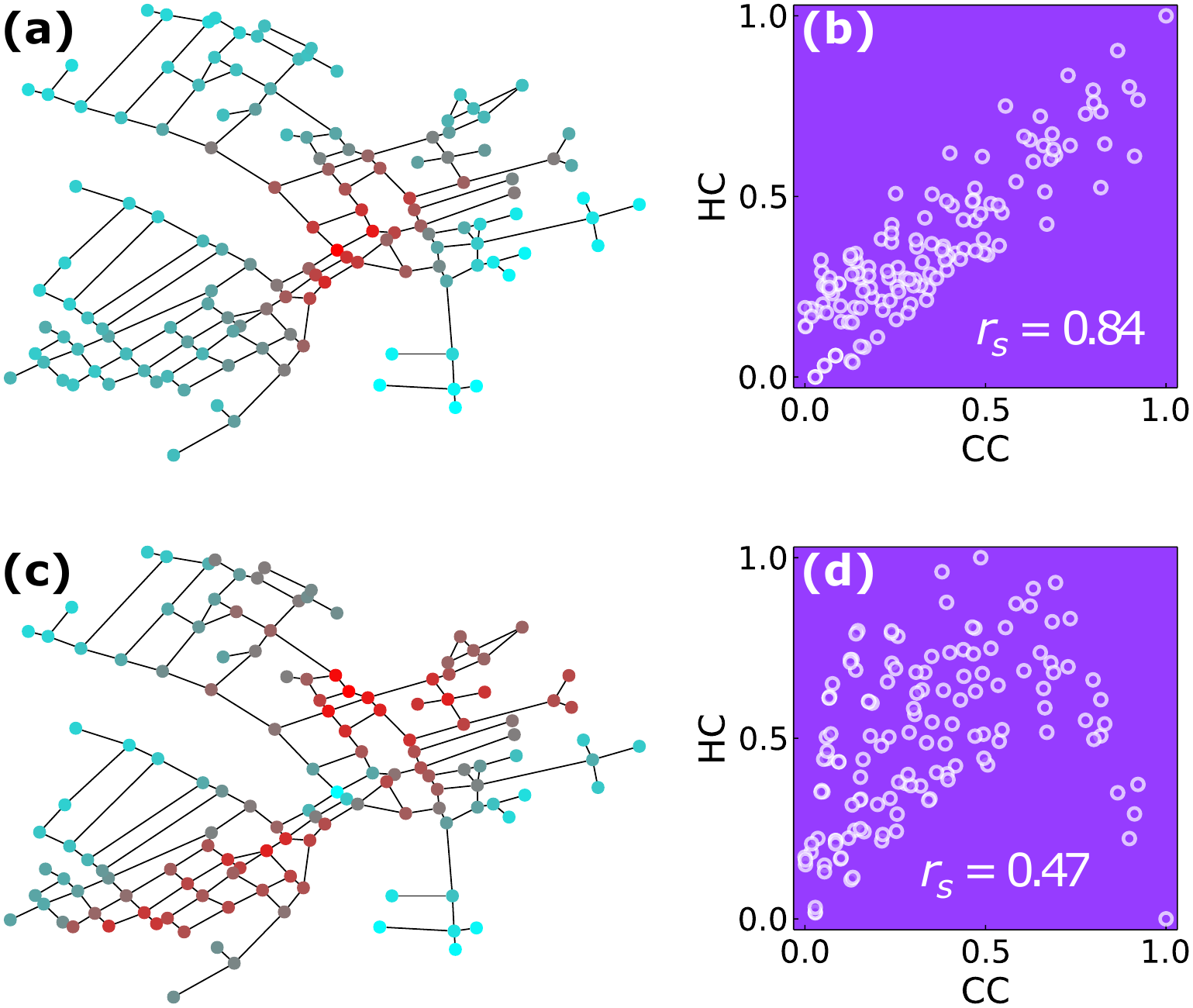}
  \caption{Hotelling centrality (HC) for the street network of Venice and its comparison with the closeness centrality (CC). Both centralities are normalized between $0$ and $1$. 
  Panels (a), (b) refer to the OS dynamics, while panels (c), (d) to the BR dynamics. The Spearman's correlation coefficient $r_s$ between the two quantities in the scatter plots is also reported. 
  }
\label{fig:hotelling_centrality}
\end{figure}

Finally, we show that the dynamics of our spatial competition model can be used to define a new 
measure of centrality~\cite{complexbook_vito}. We introduce the Hotelling centrality (HC) of a graph node  
as the average payoff earned by a seller placed on that node, where the  
averages are taken over all the possible positions of the other seller. 
Hence, nodes with highest HC are the best positions that a seller can occupy to maximize its earning, 
when the seller has no information on the position of the other seller. 
Interestingly, in the case of a chain we observe that the two nodes with the highest HC under a BR dynamics are $i = 0.25 N$ and $j = 0.75 N$ (see SM).
These positions, when occupied by the two sellers, minimize the average delivered cost for the buyers. Namely, the positions guaranteeing the average highest payoff to a seller are also the ones minimizing the transportation cost of the buyers. 
Fig.~\ref{fig:hotelling_centrality} illustrates how the HC works in a  real-world market, such as the street network of Venice (the area around Rialto bridge)~\cite{crucitti2006,cardillo2006}, and compares the HC with a standard measure of centrality, namely the closeness centrality (CC). Panels (a), (c) show the spatial distribution of HC on the map of the city, with the nodes with the highest values of centrality coloured in red. 
The scatter plots (HC, CC) in panels (b), (d) indicate that, while HC is strongly correlated to CC in the case of the OS dynamics, the two measures can give different results when instead a BR dynamics is adopted. In the latter 
case, we observe that the relation between the HC of a node and its position on the graph is highly non-trivial. E.g. the red nodes with the highest HC in panel (c) are different from those with the highest CC (i.e. the nodes 
minimizing the average distance from the other nodes), which are located in a single area around the Rialto bridge (the graph barycenter~\cite{west2001introduction}). They define 
instead two different centers (the two red areas) in the graph, one on each sides of the bridge, while the bridge has low HC (colored in cyan).  
\paragraph*{Conclusions}
Understanding how the actual spatial structure of a market  
impacts the price dynamics is of fundamental importance to design effective policies and to optimize the market both for sellers and consumers.
Our dynamical model of spatial competition, on the one hand reveals that the structure of real-world markets further amplifies the competition in price dynamics, on the other hand uses such dynamics to characterize the market topology. 
The model can also be easily generalized in different directions, e.g. considering more than two sellers, allowing both positions and prices to change at the same time, and including non-linear costs of transportation~\cite{daspremont1979, economides1993, brenner2005}.
Since price competition is a fundamental mechanism of real-world markets, 
we hope our work can open new research avenues 
with tangible practical implications.

\bibliographystyle{apsrev4-2}
\bibliography{biblio}

\cleardoublepage
\newpage
\onecolumngrid

\setcounter{figure}{0}
\setcounter{table}{0}
\setcounter{equation}{0}
\makeatletter
\renewcommand{\thefigure}{S\arabic{figure}}
\renewcommand{\theequation}{S\arabic{equation}}
\renewcommand{\thetable}{S\arabic{table}}

\setcounter{secnumdepth}{2} 

\section*{\large{Supplemental Material: Model of spatial competition on discrete markets}}
\normalsize
\vspace*{0.2 cm}

\section{Analysis of the classical Hotelling solutions on a chain}

\begin{figure}[htp]
    \includegraphics[width=0.8\textwidth]{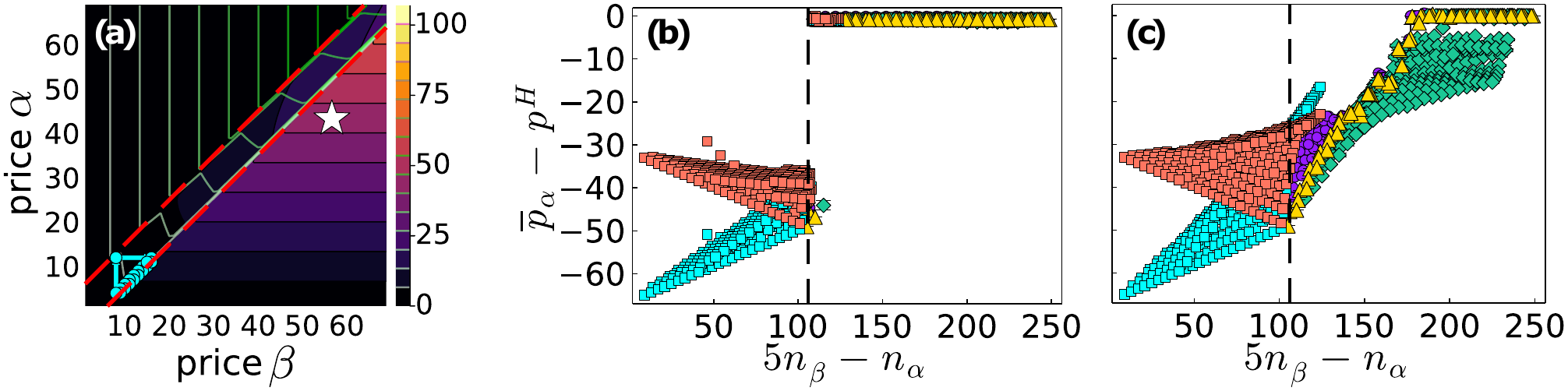}
  \caption{Comparison of the analytical predictions with the simulations results for the price dynamics on a chain of $N = 50$ nodes.
  (a) The heatmap shows the payoff of seller $\alpha$ while the payoff's level curves of seller $\beta$ are represented in green-magenta. The dashed red lines delimit the region where the price dynamics is confined to, i.e. where $|p_{\alpha} - p_{\beta}| \leq n_{\beta} - n_{\alpha}$. The white star represents the classical Hotelling equilibrium prices $(p_{\alpha}^H, p_{\beta}^H)$.
    As shown in this example for sellers positions $n_{\alpha} = 13$ and $n_{\beta}=18$, when the classical Hotelling equilibrium prices, i.e. the white star, are outside the allowed domain for the price dynamics, the price trajectory (in cyan in the plot) cannot converge to them (in this case we show the trajectory under a Best Response price update rule).
  (b), (c) Difference between the equilibrium price averaged over time $\overline{p}_{\alpha}$  and the Nash equilibrium price predicted by the classical Hotelling model as a function of $5n_{\beta} - n_{\alpha}$, respectively for One-Step (OS) and Best Response (BR) dynamics. The dashed black lines represent the sufficient condition for not having convergence to the classical Hotelling prices, given by Eq.~\eqref{eq:cond1b}. 
  The points are colored according to the sellers positions. In particular denoting the chain's middle point $x_c = 25.5$ and the sellers centre of mass $n_c = (n_{\alpha} + n_{\beta})/2$, the orange and cyan squares are for positions $(n_{\alpha}, n_{\beta})$ such that $n_{\alpha} < n_{\beta} < x_c$ and $n_{\beta} < n_{\alpha} < x_c$ respectively, while the purple circles, green diamonds and yellow triangles are for $n_{\alpha} < x_c < n_{\beta}$ with $n_c < x_c - 0.5$, $n_c > x_c + 0.5$ and $|n_c - x_c| \leq 0.5$ respectively.}
\label{fig:onchain_analytical}
\end{figure}

Let us consider a chain of $N$ nodes where the nodes are labeled with integer numbers in increasing order from $1$ to $N$. We assume that the sellers $\alpha$ and $\beta$ occupy respectively the positions (nodes) $n_{\alpha}$, $n_{\beta}$ such that $n_{\alpha}<n_{\beta}$ (being the sellers indistinguishable, the results can be readily extended to the symmetric case $n_{\beta}<n_{\alpha}$ by swapping the indices $\alpha$ and $\beta$). 
Moreover, because of the chain symmetry we can focus in our analysis only on the positions $n_{\alpha}, n_{\beta}$ such that $n_{\alpha} < (N + 1) / 2$. All the results can be easily extended to the symmetric case $n_{\alpha} > (N+1)/2$.
We start noticing that for fixed sellers positions $n_{\alpha}$, $n_{\beta}$, the sellers payoffs as a function of prices are discontinuous in
\begin{align}
p_{\alpha} &= p_{\beta} - (n_{\beta} - n_{\alpha}) 
\\
p_{\alpha} &= p_{\beta} + (n_{\beta} - n_{\alpha})
\end{align}
In fact, when the absolute value of the price difference exceeds the distance between the two sellers, i.e. if
\begin{equation}
    |p_{\alpha} - p_{\beta}| \geq n_{\beta} - n_{\alpha}
\end{equation}
the payoff of the seller with the higher price suddenly drops to $0$.
Since the sellers are rational, they will never adopt prices such that $|p_{\alpha} - p_{\beta}| \geq n_{\beta} - n_{\alpha}$ during the price dynamics (or analogously such that $|p_{\alpha} - p_{\beta}| > n_{\beta} - n_{\alpha} - 1$, since the prices are discrete). That is, if we look at Fig.~\ref{fig:onchain_analytical}a the dynamics will be always confined between the two red lines representing $|p_{\alpha} - p_{\beta}| = n_{\beta} - n_{\alpha}$.
This implies that if the classical Hotelling model's equilibrium
\begin{align}
    p_{\alpha}^H = N + \frac{\left( a - b \right)}{3} \label{eq:hp_alpha}
    \\
    p_{\beta}^H = N - \frac{\left( a - b \right)}{3} \label{eq:hp_beta}
\end{align}
where $a = n_{\alpha} - 1$ and $b = N - n_{\beta}$, are such that
\begin{equation} \label{eq:cond_extended}
    |p_{\alpha}^H - p_{\beta}^H| =  \frac{2}{3}|a - b| = \frac{2}{3}|n_{\alpha} + n_{\beta} - N - 1| \geq n_{\beta} - n_{\alpha}
\end{equation}
then the price dynamics cannot converge to them, because outside the region of prices involved in the prices dynamics. 
We now investigate for which positions $n_{\alpha}$, $n_{\beta}$ this happens.
Eq.~\eqref{eq:cond_extended} gives us two inequalities:
\begin{align}
    n_{\beta} &\leq 5n_{\alpha} - 2(N+1) \textbf{, for } n_{\alpha} + n_{\beta} > N + 1 \label{eq:cond1a}
    \\
    n_{\alpha} &\geq 5n_{\beta} - 2(N+1) \textbf{, for } n_{\alpha} + n_{\beta} < N + 1 \label{eq:cond1b}
\end{align}
That is
\begin{align} 
    N+1 - n_{\alpha}< n_{\beta} &\leq 5n_{\alpha} - 2(N+1) 
    \\
     5n_{\beta} - 2(N+1) \leq n_{\alpha} &< N + 1 - n_{\beta} \label{eq:cond_complete}
\end{align}
However, looking at the first of these inequalities we notice that in order to have $N+1 - n_{\alpha} < 5n_{\alpha} - 2(N+1) $ we need $n_{\alpha} > (N + 1)/2$. Since in our analysis we are considering $n_{\alpha} < (N+1)/2$, we can focus exclusively on Eq.~\eqref{eq:cond_complete}.
By solving $5n_{\beta} - 2(N+1) < N + 1 - n_{\beta}$, we found the necessary condition that must be satisfied in order to have a non empty set of $n_{\alpha}$ values satisfying Eq.~\eqref{eq:cond_complete}:
\begin{equation}
    n_{\beta} < \frac{N+1}{2}
\end{equation}
That is, all the couples of positions $(n_{\alpha}, n_{\beta})$ with $n_{\alpha} < n_{\beta}$ and $n_{\alpha} < (N+1)/2$ such that $\exists n_{\alpha}$ which satisfies Eq.~\eqref{eq:cond_complete} are among the positions where $n_{\beta} < \frac{N+1}{2}$, i.e. when both sellers are on the same side of the chain (that is, among the orange and cyan points in our plots).
We can also found a sufficient condition for Eq.~\eqref{eq:cond_complete}, by looking when
$5n_{\beta} - 2(N + 1) < 0$ (in this case all the values of $n_{\alpha}$ automatically satisfy Eq.~\eqref{eq:cond_complete}, since $0<n_{\alpha}$ by definition). In this way we find the sufficient condition:
\begin{equation} \label{eq:cond_sufficient}
    n_{\beta} < \frac{2}{5}(N + 1)
\end{equation}
If instead we are interested in the symmetric case where $n_{\alpha} > (N + 1) / 2$, to obtain the sufficient condition equivalent to Eq.~\eqref{eq:cond_sufficient} we just need to swap the index $\beta$ with $\alpha$, the $<$ sign with $>$ and subtract $N + 1$ to the right-hand side of the inequalities, obtaining: $n_{\alpha} > \frac{3}{5}(N+1)$ (and where as usual $n_{\alpha} < n_{\beta}$).
It is crucial to notice that Eq.~\eqref{eq:cond_complete} tells us the condition that must be satisfied for the Hotelling solution to be outside of the allowed domain of the price dynamics, but this is only a sufficient condition for not observing the classical Hotelling equilibrium. That is, it does not imply that if the Hotelling solutions are instead inside the domain interested by the price dynamics then the prices will converge to them.
In particular, with the numerical simulations we found  that for the One-Step dynamics the condition is actually necessary and sufficient for not observing the classical Hotelling solutions, i.e. all and only the positions for which the actual model price dynamics does not converge to the classical Hotelling solution satisfy Eq.~\eqref{eq:cond_complete}, as shown in Fig.~\ref{fig:onchain_analytical}b. Instead, in Fig.~\ref{fig:onchain_analytical}c we see that for the Best Response dynamics the price dynamics does not converge to the classical Hotelling solutions even for positions which do not satisfy Eq.~\eqref{eq:cond_complete}.

\section{One-step price dynamics convergence to the classical Hotelling equilibrium}

We consider the case where the sellers are allowed to change their price by $\pm 1$ at each time step (i.e. when sellers use the One-Step, OS, price update rule) and we prove that the price dynamics converges to the classical Hotelling equilibrium (when it can, see previous SM section).
Let us consider two sellers $\alpha$ and $\beta$ respectively on node $n_{\alpha}$ and $n_{\beta}$ of a chain of $N$ nodes, where the nodes are labelled with integers numbers $ 1 \leq n \leq N$ in increasing order.
We define the \emph{indifference distance} $d_{in}$ as the distance between seller $\alpha$ and the node with equal delivered price from the two sellers:
\begin{equation} \label{eq:indif_dist}
    d_{in} = \frac{d_{\alpha \beta}}{2} + \frac{p_{\beta} - p_{\alpha}}{2}
\end{equation}
where $d_{\alpha \beta} = n_{\beta} - n_{\alpha}$ is the distance between the two sellers and $p_{\alpha}$, $p_{\beta}$ are their prices.
Fixed the sellers position (i.e. their distance), the buyers flux $\phi$ (i.e. the share of buyers market) attracted by each seller at time $t$ is a function of $d_{in}^t$ the current indifference distance, i.e. of their current prices:
\begin{align}
    \phi_{\alpha}^t &=  a + d_{in}^t + \frac{1}{2} = n_{\alpha} + d_{in}^t - \frac{1}{2}
    \\
    \phi_{\beta}^t &= b + + d_{in}^t + \frac{1}{2} = N - n_{\beta} + d_{\alpha \beta} - d_{in}^t + \frac{1}{2}
\end{align}
where $a=n_{\alpha} - 1$ and $b = N - n_{\beta}$. It is worth to point out that these equations for the buyers fluxes apply only for $|p_{\beta} - p_{\alpha}| < n_{\beta} - n_{\alpha}$ with $n_{\alpha} < n_{\beta}$.
The payoff $\pi$ of each seller can be readily found multiplying the buyers market share by the respective prices at time $t$ $p_{\alpha}^t$ and $p_{\beta}^t$:
\begin{align}
    \pi_{\alpha}^t &= \left(n_{\alpha} + d_{in}^t - \frac{1}{2}\right)p_{\alpha}^t \label{eq:pay_t}
    \\
    \pi_{\beta}^t &= \left(N - n_{\beta} + d_{\alpha \beta} - d_{in}^t + \frac{1}{2}\right)p_{\beta}^t
\end{align}
Let us focus on seller $\alpha$ and suppose that at time $t$ it changes the price $p_{\alpha}^{t+1}=p_{\alpha}^t \pm 1$. As a consequence the indifference distance at time $t+1$ will be $d_{in}^{t+1}=d_{in}^t \mp 1/2$, since by increasing (decreasing) the price by $1$, seller $\alpha$ decreases (increases) its buyers' market share by $1/2$.
In particular, by decreasing the price the payoff at time $t+1$ becomes
\begin{equation} \label{eq:pay_decreasing}
    \pi_{\alpha,-}^{t+1} := \pi_{\alpha}^{t+1}(p_{\alpha}^t - 1) = \left( n_{\alpha} + d_{in}^{t+1} - \frac{1}{2} \right)\left( p_{\alpha}^t - 1 \right) = \left( n_{\alpha} + d_{in}^{t} + \frac{1}{2} - \frac{1}{2} \right)\left( p_{\alpha}^t - 1 \right) = \pi_{\alpha}^{t} + \frac{1}{2}p_{\alpha}^t - \left( n_{\alpha} + d_{in}^t \right)
\end{equation}
while increasing the price by $1$ the seller earns
\begin{equation} \label{eq:pay_increasing}
    \pi_{\alpha,+}^{t+1} := \pi_{\alpha}^{t+1}(p_{\alpha}^t + 1)  = \left( n_{\alpha} + d_{in}^{t+1} - \frac{1}{2} \right)\left( p_{\alpha}^t + 1 \right) = \left( n_{\alpha} + d_{in}^t - \frac{1}{2} - \frac{1}{2} \right)\left( p_{\alpha}^t + 1 \right) = \pi_{\alpha}^{t} - \frac{1}{2}p_{\alpha}^t + \left( n_{\alpha} + d_{in}^t - 1 \right)
\end{equation}
We now compare Eqs.~\eqref{eq:pay_decreasing},~\eqref{eq:pay_increasing}
to Eq.~\eqref{eq:pay_t} to see when it is convenient for seller $\alpha$ to keep the price fixed and when instead is more convenient to change it, i.e. when:
\begin{align}
    \pi_{\alpha}^{t} &> \pi_{\alpha,-}^{t+1} = \pi_{\alpha}^t + \frac{1}{2}p_{\alpha}^t-\left( n_{\alpha} + d_{in}^t \right)
    \\
    \pi_{\alpha}^{t} &> \pi_{\alpha,+}^{t+1} = \pi_{\alpha}^t - \frac{1}{2}p_{\alpha}^t + \left( n_{\alpha} + d_{in}^t -1\right)
\end{align}
That is 
\begin{align}
    \frac{1}{2}p_{\alpha}^t - \left( n_{\alpha} + d_{in}^t \right) < 0
    \\
     - \frac{1}{2}p_{\alpha}^t + \left( n_{\alpha} + d_{in}^t -1\right) < 0
\end{align}
By substituting Eq.~\eqref{eq:indif_dist} we obtain
\begin{align}
    2 p_{\alpha}^t - p_{\beta}^t &< d_{\alpha \beta} + 2 n_{\alpha} 
    \\
    2 p_{\alpha}^t - p_{\beta}^t &> d_{\alpha \beta} + 2(n_{\alpha} -1 )
\end{align}
These two inequalities combined together give us the condition for which it is not convenient for the seller $\alpha$ to change the price, i.e. when the price update dynamics stops:
\begin{equation} \label{eq:diseq_stop_alfa}
    d_{\alpha \beta} + 2(n_{\alpha} -1 ) < 2 p_{\alpha}^t - p_{\beta}^t < d_{\alpha \beta} + 2 n_{\alpha}
\end{equation}
And since the price are discrete and the minimum price variation is $\Delta p \pm 1$, this implies:
\begin{equation} \label{eq:cond_stop_alfa}
    2 p_{\alpha}^t - p_{\beta}^t = d_{\alpha \beta} + 2n_{\alpha} - 1
\end{equation}
For seller $\beta$ we can find an equivalent condition. In particular, by taking advantage of the symmetry of the chain it is sufficient to swap in Eq.~\eqref{eq:diseq_stop_alfa} indices $\alpha$ and $\beta$ and to replace $n_{\alpha} - 1$ with $N - n_{\beta}$:
\begin{equation}
     d_{\alpha \beta} + 2(N - n_{\beta}) < 2p_{\beta} - p_{\alpha} < d_{\alpha \beta} + 2(N - n_{\beta} + 1)
\end{equation}
That gives us:
\begin{equation} \label{eq:cond_stop_beta}
    2 p_{\beta}^t - p_{\alpha}^t = d_{\alpha \beta} + 2(N - n_{\beta}) + 1
\end{equation}
From Eq.~\eqref{eq:cond_stop_alfa} we obtain $p_{\beta} = 2 p_{\alpha} - d_{\alpha \beta} -2 n_{\alpha} + 1$, and by substituting it and $d_{\alpha \beta} = n_{\beta} - n_{\alpha}$ in Eq.~\eqref{eq:cond_stop_beta} we finally arrive to the stationary price for seller $\alpha$ under OS price update:
\begin{equation}
    p_{\alpha}^* = N + \frac{(n - 1) - (N - n_{\beta})}{3} = N + \frac{a - b}{3}
\end{equation}
since by definition $a = n_{\alpha} - 1$ and $b = N - {n_{\beta}}$.
Analogously we can obtain $p_{\alpha}$ from Eq.~\eqref{eq:cond_stop_beta}, and by replacing in Eq.~\eqref{eq:cond_stop_alfa} we find:
\begin{equation}
    p_{\beta}^* = N - \frac{(n - 1) - (N - n_{\beta})}{3} = N - \frac{a - b}{3}
\end{equation}
Therefore the stationary prices under the OS dynamics coincides with the classical Hotelling equilibrium prices, at least for sellers positions such that the price dynamics can converge to the classical Hotelling solutions (i.e. when Eq.~\eqref{eq:cond_complete} holds).

\section{Edgeworth cycles for bounded information and sellers on the same node}

We assume that a fraction $1-w$ (where $0<w<1$) of the buyers has bounded information regarding the sellers positions and prices. 
For this analytical treatment we assume that the two sellers $\alpha$ and $\beta$ are on the same node. We recall from the main text that the fraction $w$ of informed buyers will buy from the cheaper seller while $1-w$ will buy with equal probability from the two sellers on the same node, as long as the price difference $\Delta p$ is below the threshold $\Delta p_{T} > 0$. We can write the normalized (i.e. divided by the cardinality $N$ of the network) fluxes of buyers attracted by the two sellers as a function of the price difference.
If $p_{\alpha} = p_{\beta}$, each seller attracts half of the total market $\phi_{\alpha} = \phi_{\beta} = \frac{1}{2}$. 
If $ 0 < p_{\beta} - p_{\alpha} \leq \Delta p_{T}$
\begin{align}
\phi_{\alpha} &= w + \frac{1 - w}{2} = \frac{1 + w}{2}
\\
\phi_{\beta} &= \frac{1 - w}{2}
\end{align}
while if $p_{\beta} - p_{\alpha} > \Delta p_{T}$, $\alpha$ get all the buyers, i.e. $\phi_{\alpha} =1$ and $\phi_{\beta} = 0$.
The sellers payoffs are readily found by multiplying the fluxes of buyers for the respective prices $p_{\alpha}$ and $p_{\beta}$.
It is worth to notice that the case $p_{\beta} - p_{\alpha} > \Delta p_{T}$ clearly will never occur if sellers are rational, since by decreasing the price difference below the critical threshold seller $\beta$ will earn some positive payoff instead than a null one.
Hence we can focus on the case where the price difference is below $\Delta p_{T}$. In particular we will compare the sellers payoffs to understand when for a seller is convenient to keep the same price of the competitor and when instead it is convenient to raise its price above, or decrease it below, the competitor price.
We focus on seller $\alpha$, considering the price of $\beta$ fixed. Being the sellers identical (they are both on the same node) the same analysis applies if we swap sellers' indices.
We first notice that it is never convenient for the seller $\alpha$ to lower the price more than the minimum discrete price (i.e. a unit of price) below the competitor price, since its share of market is equal to $\phi_{\alpha} =  \frac{1 + w}{2}$ for all prices below $p_{\beta}$, while instead the payoff earned $\pi_{\alpha} = \phi_{\alpha}p_{\alpha}$ increases linearly with $p_{\alpha}$. That is, if a rational seller $\alpha$ decreases its price below $p_{\beta}$, it will adopt $p_{\alpha} = p_{\beta} - 1$.
For the same reason, since increasing $p_{\alpha}$ above $p_{\beta}$ guarantees a buyers market $\phi_{\alpha} = \frac{1 - w}{2}$ for every price $p_{\alpha} > p_{\beta}$ (as long as $ 0 < p_{\alpha} - p_{\beta} \leq \Delta p_{T}$), if it is convenient for $\alpha$ to increase the price then the best payoff it can obtain is for the maximum price difference below the threshold, i.e. for $p_{\alpha} = p_{\beta} + \Delta p_{T}$.

\bigskip

$\bullet$ \textbf{Same price or to lower?}

That is, we have to check when:
\begin{equation} \label{eq:same_or_lower}
    \frac{p_{\beta}}{2} > \left( \frac{1 + w}{2} \right)\left(p_{\beta}-1\right)
\end{equation}
where on the LHS we have the payoff of seller $\alpha$ for adopting the same price of seller $\beta$, on the RHS the payoff for a price $p_{\alpha} = p_{\beta} - 1$.
With some basic manipulation we obtain:
\begin{equation} \label{eq:same_or_lower_w}
    1-w > \frac{p_{\beta} - 2}{p_{\beta} - 1}
\end{equation}
Or analogously, focusing on the price as a function of $w$, we obtain:
\begin{equation} \label{eq:same_or_lowe_p}
    p_{\beta} < \frac{1+w}{w}
\end{equation}
Since the fraction of uninformed buyers is strictly greater than zero, Eq.~\eqref{eq:same_or_lower_w} implies that if $p_{\beta} = 2$ there is no value of $1-w > 0$ for which is convenient to lower the price further (i.e. the minimum possible price is $p_{\beta} = 2$).

\bigskip

$\bullet$ \textbf{Same price or to rise?}

In this case we have to verify when
\begin{equation} \label{eq:same_or_rise}
    \frac{p_{\beta}}{2} > \left( \frac{1 - w}{2} \right)\left(p_{\beta} + \Delta p_{T}\right)
\end{equation}
where on the LHS there is the payoff of seller $\alpha$ for adopting the same price of seller $\beta$, on the RHS the payoff for a price $p_{\alpha} = p_{\beta} + \Delta p_{T}$.
This gives us:
\begin{equation} \label{eq:same_or_rise_w}
    1-w < \frac{p_{\beta}}{p_{\beta} + \Delta p_{T}}
\end{equation}
Or if we focus on the price as a function of $w$:
\begin{equation} \label{eq:same_or_rise_p}
    p_{\beta} > \frac{(1-w)\Delta p_{T}}{w}
\end{equation}

\bigskip

$\bullet$ \textbf{To lower or to rise?}

In this case we have to verify when
\begin{equation} \label{eq:lower_or_rise}
    \frac{1+w}{2}(p_{\beta} -1) > \left( \frac{1 - w}{2} \right)\left(p_{\beta} + \Delta p_{T}\right)
\end{equation}
where on the LHS we have the payoff of seller $\alpha$ for adopting a price $p_{\alpha} = p_{\beta} - 1$, on the RHS the payoff for a price $p_{\alpha} = p_{\beta} + \Delta p_{T}$.
This gives us:
\begin{equation} \label{eq:lower_or_rise_w}
    1-w < \frac{2(p_{\beta}-1)}{2p_{\beta} + \Delta p_{T} - 1}
\end{equation} \label{eq:lower_or_rise_p}
Or if we focus on the price:
\begin{equation}
    p_{\beta} > \frac{(1-w)\Delta p_{T} + 1}{2w} + \frac{1}{2} =: p_{th}
\end{equation}
This means that as long as the price of the competitor is higher than this critical threshold $p_{th}$ the sellers will undercut the competitor price by one. Once this critical threshold is reached then one of the seller (e.g. seller $\alpha$) will increase its price to the maximum allowed, i.e. to $p_{\alpha} = p_{th} + \Delta p_{T}$.
However, from Eq.~\eqref{eq:same_or_lower_w} we know that the minimum price reachable undercutting the competitor is $p = 2$, otherwise for all $1-w>0$ it is more convenient for the sellers to charge the same price than to reduce the price further.
By substituting $p_{min} = 2$ in Eq.~\eqref{eq:lower_or_rise_w} we found that if
\begin{equation}
    (1-w)<\frac{2(p_{min}-1)}{2 p_{min}+\Delta p_{T} - 1 } = \frac{2}{3+\Delta p_{T}} =:(1-w)'
\end{equation}
then $p_{th} < 2$.
That is, for $(1-w)<(1-w)'$ the prices undercutting process will stop when both sellers reach the same price $p_{min} = 2$.
At this point we have to check the condition given by Eq.~\eqref{eq:same_or_rise}, to see if for a seller it is more convenient to stay at the same price of the competitor $p_{min}=2$ or to rise its price. Substituting $p_{min} = 2$ in Eq.~\eqref{eq:same_or_rise_w} we see that, as long as 
\begin{equation}
  1-w > \frac{2}{2 + \Delta p_{T}} =: (1 - w)''  
\end{equation}
it is more convenient to rise the price. Otherwise, since $(1-w)'< (1-w)''$, the price dynamics remains stuck in $p_{\alpha} = p_{\beta} = p_{min} = 2$.
\begin{figure}[htp]
    \includegraphics[width=0.5\textwidth]{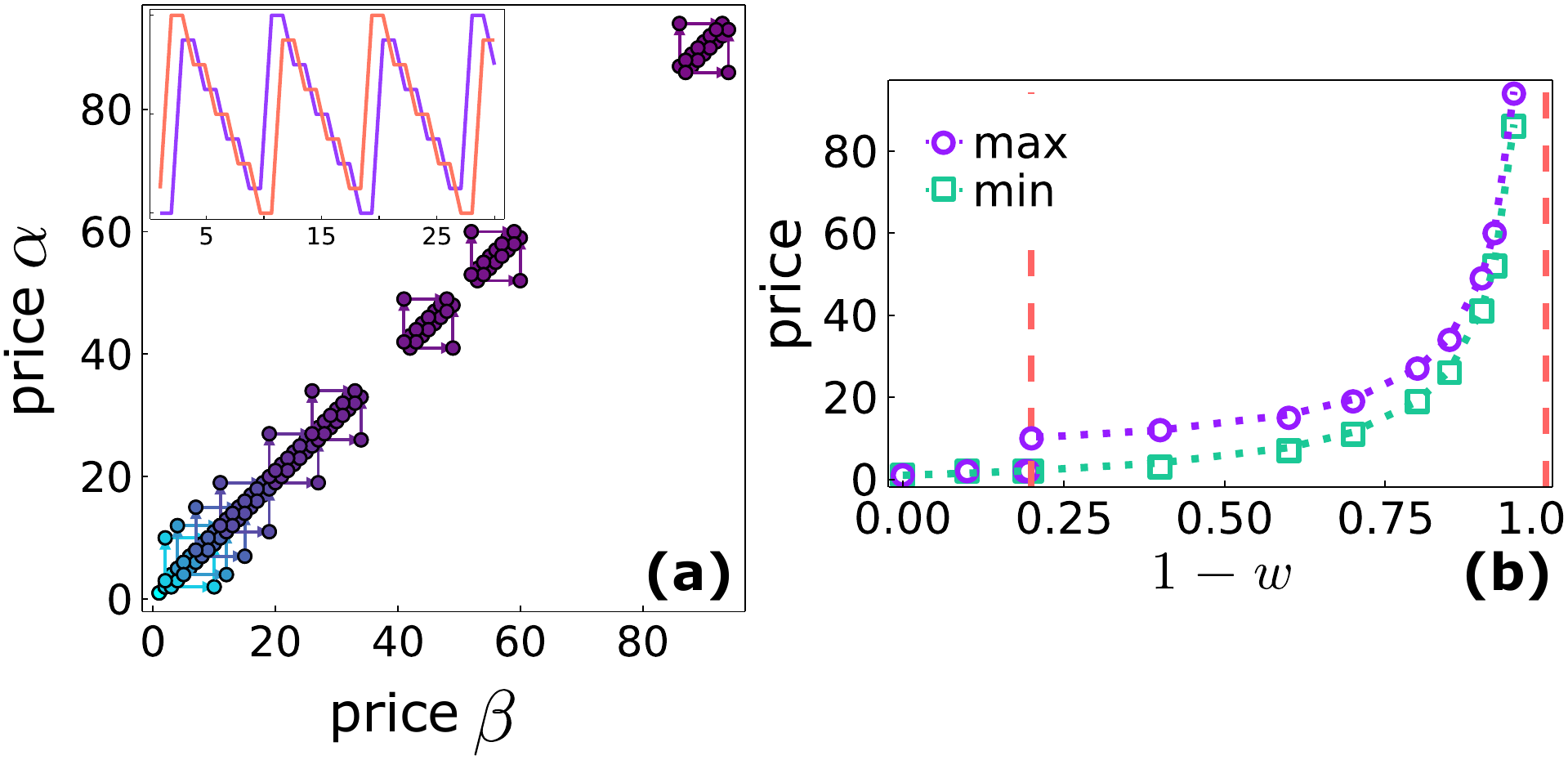}
  \caption{(a), (b) Edgeworth cycles as a function of $0 < (1-w) < 1$, i.e. for bounded buyers' rationality, when both sellers are on the same node. In this case the price cycles are identical for the two sellers (being the sellers indistinguishable and on the same node). (b) We see  that the numerical results (colored dots) are in perfect agreement with the analytical predictions. In particular the dashed vertical lines delimit the domain in $1-w$ given by Eq.~\eqref{eq:domain_cycle} where cycles exist, while the green and purple dotted lines are respectively the minimum and maximum prices given by Eqs.~\eqref{eq:minimumcycle},~\eqref{eq:maximumcycle}.}
\label{fig:wrw_characterization}
\end{figure}
Hence, Edgeworth cycles exist for 
\begin{equation}
    \frac{2}{2 + \Delta p_{T}} < 1-w < 1 \label{eq:domain_cycle}
\end{equation}
while the minimum and maximum prices in each cycle are given by:
\begin{align}
    p_{min} &= \frac{(1-w)\Delta p_{T} + 1}{2w} + \frac{1}{2} \label{eq:minimumcycle}
    \\
    p_{max} &= \frac{(1-w)\Delta p_{T} + 1}{2w} + \frac{1}{2} + \Delta p_{T} \label{eq:maximumcycle}
\end{align}
In Fig.~\ref{fig:wrw_characterization} we compare the theoretical predictions with the simulations results, finding a perfect agreement.

\section{Bounded information and sellers on different chain's nodes}

\begin{figure}[htp]
    \includegraphics[width=1.0\textwidth]{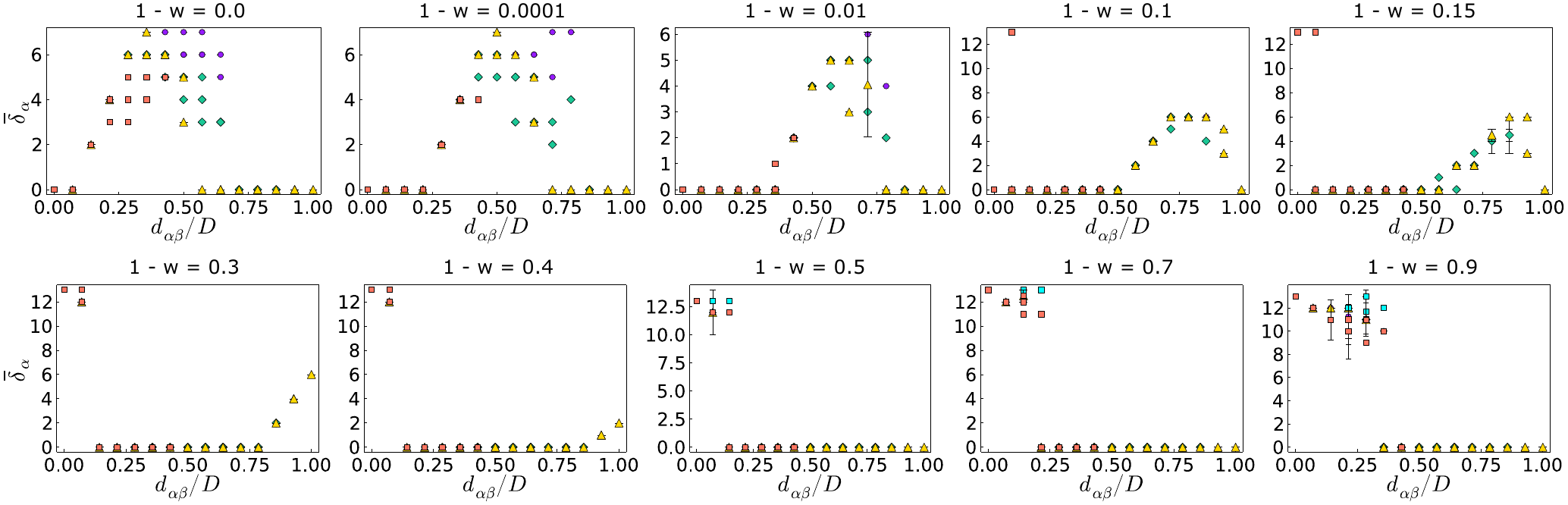}
  \caption{Numerical results for BR price dynamics. In particular the panels show the average amplitude of the price cycles $\overline{\delta}_{\alpha}$ as a function of the normalized distance between the sellers $d' = d_{\alpha \beta} / D$, where D is the chain's diameter (these results have been obtained for a chain of $N=15$ nodes). Each panel refers to a different value of $w$, the fraction of informed buyers. The points with $\delta_{\alpha}= 0$ correspond to fixed points of the price dynamics.}
\label{fig:wrw_distinctnodes_amplitude_BR}
\end{figure}

\begin{figure}[htp]
    \includegraphics[width=1.0\textwidth]{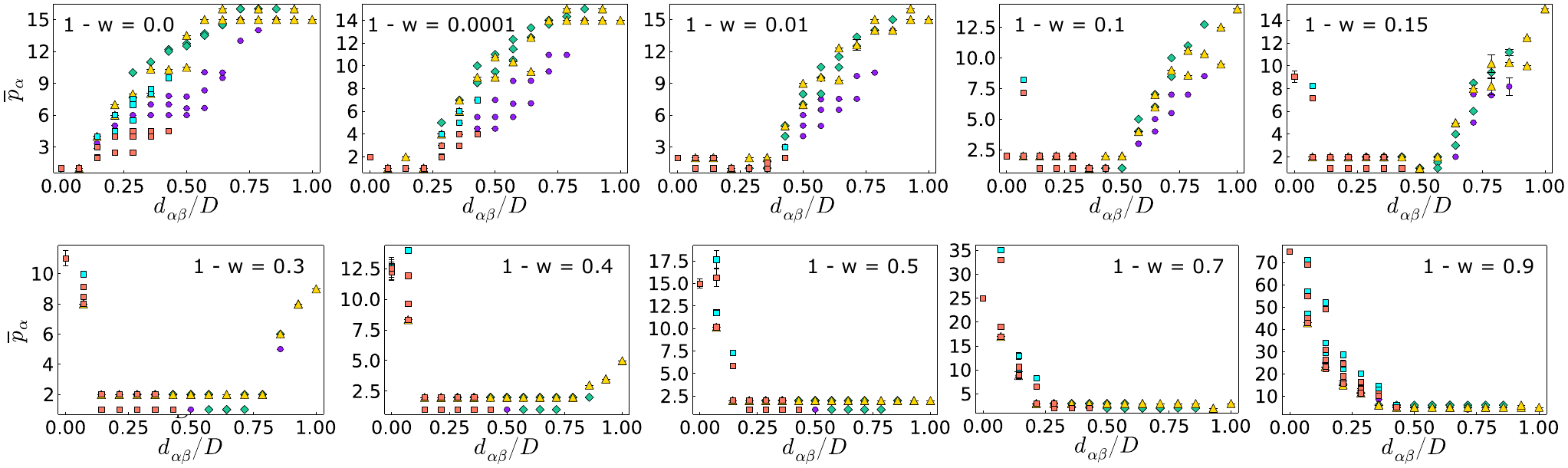}
  \caption{Numerical results for BR price dynamics. In particular we report the average price $\overline{p}_{\alpha}$ as a function of the normalized distance between the sellers $d' = d_{\alpha \beta} / D$, where D is the chain's diameter (these results have been obtained for a chain of $N=15$ nodes). Each panel refers to a different value of $w$, the fraction of informed buyers.}
\label{fig:wrw_distinctnodes_price_BR}
\end{figure}

\begin{figure}[htp]
    \includegraphics[width=1.0\textwidth]{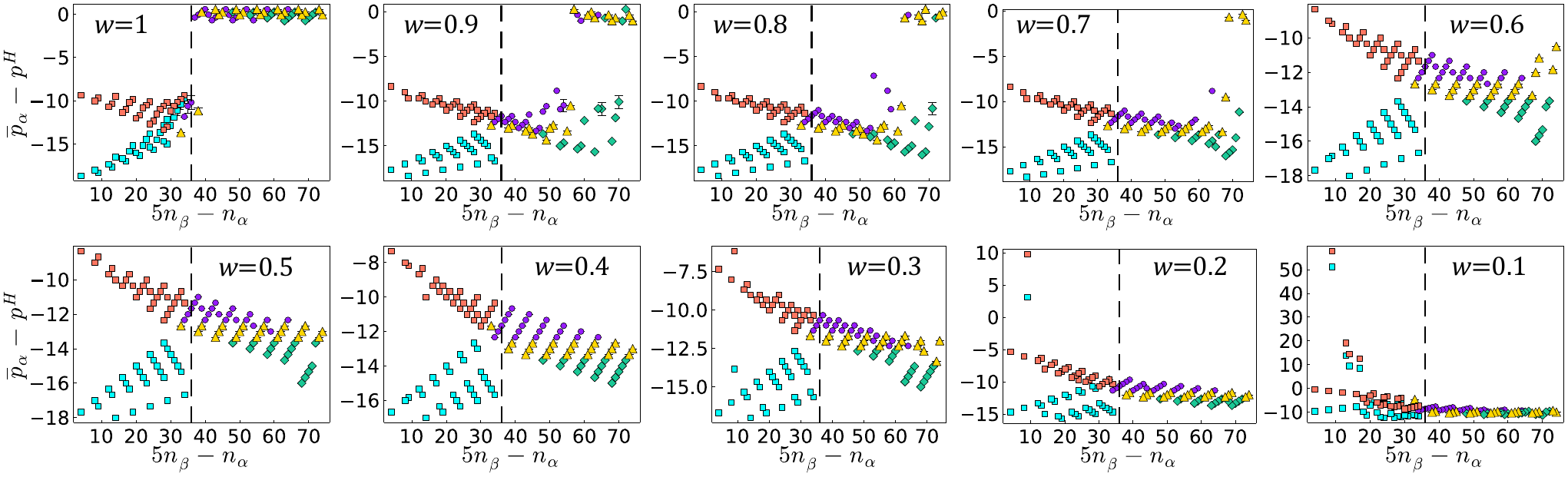}
  \caption{Difference between the average prices for OS dynamics and the classical Hotelling Nash equilibrium prices given by Eqs.~\eqref{eq:hp_alpha},~\eqref{eq:hp_beta}, as a function of $5 n_{\beta} - n_{\alpha}$ (see first section of the SM). Each panel refers to a different value of $w$, the fraction of informed buyers. The vertical dashed lines mark the sufficient condition Eq.~\eqref{eq:cond1b} for not having convergence to the classical Hotelling equilibrum prices.}
\label{fig:wrw_distinctnodes_OS}
\end{figure}

We characterize numerically the impact of $w$ over the price dynamics, both for the BR (Figs.~\ref{fig:wrw_distinctnodes_amplitude_BR},~\ref{fig:wrw_distinctnodes_price_BR}) and OS (Fig.~\ref{fig:wrw_distinctnodes_OS}) price updates, as a function of the sellers positions on a chain.
In particular, in Fig.~\ref{fig:wrw_distinctnodes_amplitude_BR} and Fig.~\ref{fig:wrw_distinctnodes_price_BR} we report respectively the average amplitude of the price cycles $\overline{\delta}_{\alpha}$ and the average price $\overline{p}_{\alpha}$ as a function of the normalized distance between the two sellers $d' = d_{\alpha \beta} / D$, where D is the chain's diameter (for these results we used a chain of $N=15$ nodes). Each panel refers to a different value of $w$, the fraction of informed buyers.
It is worth to notice that the points with $\overline{\delta}_{\alpha}= 0$ correspond to fixed points of the price dynamics, while the points with $\overline{\delta}_{\alpha} > 0$ correspond to price cycles.
As the fraction of informed buyers decreases (i.e., $1 - w$ increases), we observe that the price cycles move towards higher values of $d'$. It is interesting to notice that even very small values of $0 < 1 - w \leq 0.01$, i.e. when just a small fraction of the buyers is uninformed about sellers' prices and positions, have a noticeable impact over the price dynamics. For values of $1-w$ larger than $\approx 0.1$ we observe that price cycles appear for small value of $d'$. By further increasing $1-w$, cycles gradually disappear for large values of $d'$, while on the left of the $d'$ axis cycles extend to larger $d'$.
In Fig.~\ref{fig:wrw_distinctnodes_price_BR} we can observe that for $w>0.1$ the average prices of the cycles are typically higher than the prices for the fixed points.
Fig.~\ref{fig:wrw_distinctnodes_OS} shows the difference between the average prices for OS price dynamics and the classical Hotelling Nash equilibrium~\cite{acourseinGT} prices given by Eqs.~\eqref{eq:hp_alpha},~\eqref{eq:hp_beta}, as a function of $5 n_{\beta} - n_{\alpha}$ (see first section of the SM).
For $w=1$ we observe that the sufficient condition given by Eq.~\eqref{eq:cond1b} for not having convergence to the classical Hotelling equilibrium prices is also a necessary condition, as shown in the first section of the SM. 
As $w$ decreases we observe that more and more points (for increasing values of $5 n_{\beta} - n_{\alpha}$) diverge from the classical Hotelling solutions.
Interestingly, our numerical simulations show that, when this divergence initially occurs, the measured average price is always smaller than the one predicted by the classical Hotelling model.
Further decreasing $w$ (e.g. for $0.1 \leq w \leq 0.2$, as shown in Fig.~\ref{fig:wrw_distinctnodes_OS}) we observe that for small values of $5 n_{\beta} - n_{\alpha}$ some points start showing an increased payoff respect to the classical Hotelling prediction.
It is worth to notice that the case $w=0$, i.e. when we have just uninformed buyers, is trivial: the prices will increase indefinitely, independently from the sellers positions, since the sellers can gradually increase their prices (keeping the prices difference below the threshold $\Delta p_{T}$, introduced in main text) without any reaction from the buyers.

\section{Market competition dimension measurements}

\begin{figure}[htp]
    \includegraphics[width=1.0\textwidth]{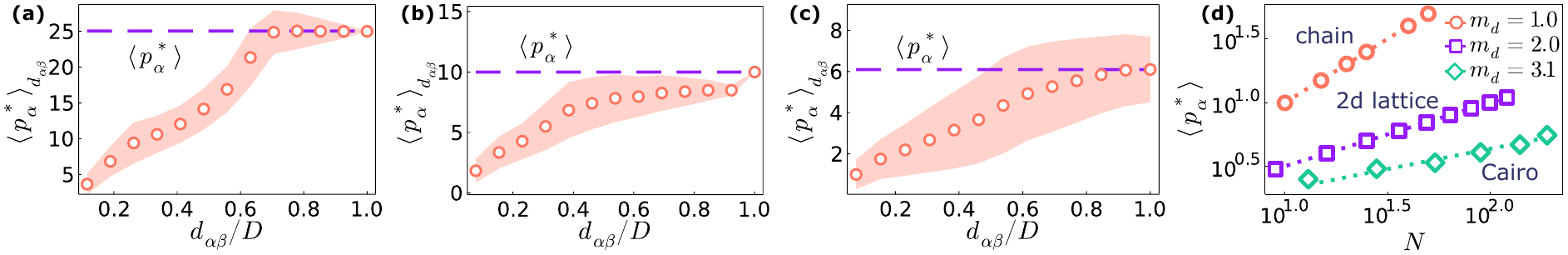}
  \caption{Maximum average equilibrium price $\langle p^*_{\alpha}\rangle$ for different network topologies and its scaling with the network order $N$. For these plots we used the BR price update rules. Panels (a), (b) and (c) refer respectively to a chain of $N=25$ nodes, a square lattice of $N=100$  nodes (i.e. a 10x10 grid) and the street networks of the city of Cairo with $N=190$ nodes (it has been obtained taking a neighbourhood of range $r=7$ of the node in the original dataset with the highest closeness centrality (for more details see next section of the SM). (d) shows the values of  $\langle p^*_{\alpha}\rangle$ for the three different networks as a function of the network order $N$ and reports the market competition dimensions $m_d$ obtained by fitting the scaling relation.}
\label{fig:md_measurements}
\end{figure}

It is trivial to prove that on a linear market, given Eqs.~\eqref{eq:hp_alpha},~\eqref{eq:hp_beta}  and $d_{\alpha \beta} = N - a - b$, the expected price $\langle p_{\alpha}^H \rangle_{d_{\alpha \beta}}$ for a distance $d_{\alpha \beta}$ is actually independent on $d_{\alpha \beta}$ and it is simply equal to $\langle p_{\alpha}^H\rangle_{d_{\alpha \beta}} = \langle p_{\alpha}^H\rangle = N$, where $\langle \cdot \rangle_{d_{\alpha \beta}}$ indicates the average over all the sellers positions at distance $d_{\alpha \beta}$.
However, as we have shown in the first section of the SM, the classical Hotelling solutions are not expected to hold for all the positions of the sellers.
In particular, on a chain market of order $N$ we observe that the expected equilibrium price is a function of the sellers' distance and it converges to a maximum value $\langle p_{\alpha}^*\rangle \sim N$ for both OS and BR dynamics.
For market structures more complex than a simple line (even a regular square lattice) finding the equilibrium prices is not trivial, if possible at all, hence we rely on numerical simulations for the characterization of the equilibrium prices.
In Fig.~\ref{fig:md_measurements} we can observe than $\langle p_{\alpha}^* \rangle$ is indeed a function of $N$ and of the market structure (i.e. the topology of the graph).
To measure the market competition dimension $m_d$ we first found the maximum of the average price $\langle p_{\alpha}^* \rangle$ as a function of the seller distance for different values of the network order $N$.
In the case of artificial networks (i.e. for chains and square lattices) we obtained networks of different order $N$ directly by generating chains and square lattices for different values of $N$.
Instead for the street networks of real-world cities we have considered the subgraphs defined by a neighbourhood of range $r$ of the node with the highest closeness centrality for different values of $r$ (for more details see the next section of the SM). Then we fitted the numerical results with the the scaling relation (see main text):
\begin{equation}
    \langle p^* \rangle \sim N^{1/m_d}
\end{equation}

\section{Market competition dimension for real-world street networks}

We report the market competition dimension $m_d$ for the street networks of real-world cities.
For this measure we used a dataset of the street networks of 19 cities from all over the world~\cite{crucitti2006,cardillo2006}.
We recall from the main text that $m_d$ is defined by
\begin{equation}
    \langle p^* \rangle = 2 \langle \pi^* \rangle \sim N^{1/m_d}
\end{equation}
and it represents how the maximum average seller's price (which is equal to two times the maximum average seller's payoff) as a function of the sellers' distance scales with $N$, the cardinality of the market's network. The averages are performed over all the possible combinations of sellers positions at a given distance $d_{\alpha \beta}$.
We performed this measure with two different methods to check the consistency of our results.
We first sampled a subgraph of the street networks: starting from the more central node in the graph, measured as the node with the highest closeness centrality, we took the subgraph defined by a neighbourhood of range $r$, for $2 \leq r \leq 7$. Obviously by increasing $r$ also $N$ the number of nodes in the subgraph increases. The choice of a maximum value of $r=7$ has two motivations. The first one is related to the dataset we used: since the street network dataset of each city represents a geographical map of finite size, in most of the cities for $r>7$ we are already hitting the edges of the finite map. Therefore further increasing $r$ (and hence $N(r)$) we would not observe how the market expands with $N$ since the expansion is bounded by the finiteness of the map. Instead, for $2 \leq r \leq 7$ we have a consistent range of $N(r)$ spanning more then one order of magnitude (from $N(r) \approx 10$ to $N(r)\approx 100$) for all cities, without reaching the edges of the street maps. The second is instead a practical reason: since we have to run simulations of the price dynamics for all the possible combinations of positions of the two sellers, the possible combinations increase roughly as $N^2$ and so increasing further $N(r)$ would have brought to impractical simulations' length. 
For each subgraph of size $N(r)$ we measured $\langle p^* \rangle(N) = 2 \langle \pi^* \rangle(N)$ as two times the maximum of the average payoff as a function of the sellers distance. Given a graph of size $N$, to find this maximum we first averaged the payoff over the last 200 time steps of the price dynamics after waiting a thermalization time of 100 time steps, for each couple of sellers positions. Then we averaged over all the possible combinations of sellers positions at a given distance $d_{\alpha \beta}$ and we took the maximum over $d_{\alpha \beta}$ of the resulting average payoff.
It is worth to notice that for these results we used the BR price dynamics, however the maximum average payoff (i.e. price) does not depend on the specific dynamics (i.e. BR or OS) but only on the specific  network. Hence for each city we fitted the values $\langle p^* \rangle$ obtained for the different $N$  by  $\langle p^* \rangle\sim N(r)^{1/m_d}$ to find the market competition dimension $m_d$.
The second method that we used, that we will refer to as the \emph{brute force} method, is to directly obtain  $m_d$ for the largest sample of size $N_{max}$ (i.e. the one for $r=7$) from  $\langle p^* \rangle = N_{max}^{1/m_d}$, as $m_d= \log(N_{max}) / \log(\langle p^* \rangle)$.
We report in Table~\ref{tab:cities_md} the fitted values of $m_d$.
\begin{table}[h!]
    \setlength{\extrarowheight}{4pt}
    \begin{tabular}{c|c|c}
    \cline{1-3}
    \multicolumn{1}{c|}{\textbf{city}} & \multicolumn{1}{c|}{\textbf{$N(r)^{1/m_d}$ fit}} & \multicolumn{1}{c}{\textbf{$N_{max}^{1/m_d}$ fit}} 
    \\\cline{1-3}
     Ahmedabad  & $2.6 \pm 0.1$ & $2.5 \pm 0.1$
     \\
     Barcelona  & $2.29 \pm 0.03$ & $2.23 \pm 0.03$
     \\
     Bologna  & $2.41 \pm 0.09$ & $2.3 \pm 0.1$
     \\
      Brasilia & $2.44 \pm 0.03$ & $2.45 \pm 0.04$
      \\
     Cairo  & $3.14 \pm 0.04$ & $3.06 \pm 0.09$
     \\
     Irvine  & $2.4 \pm 0.1$ & $2.3 \pm 0.1$
     \\
      London & $2.48 \pm 0.07$ & $2.46 \pm 0.05$
      \\
      Los Angeles  & $2.6 \pm 0.1$ & $2.4 \pm 0.1$
     \\
     New Delhi  & $2.64 \pm 0.03$ & $2.55 \pm 0.07$
     \\
      New York & $2.28 \pm 0.02$ & $2.26 \pm 0.04$
      \\
      Paris  & $2.5 \pm 0.1$ & $2.4 \pm 0.1$
     \\
     Richmond  & $2.7 \pm 0.1$ & $2.6 \pm 0.1$
     \\
      San Francisco & $2.2 \pm 0.1$ & $2.1 \pm 0.1$
      \\
      Savannah  & $2.53 \pm 0.04$ & $2.53 \pm 0.04$
     \\
     Seoul  & $2.4 \pm 0.1$ & $2.3 \pm 0.1$
     \\
      Venice & $2.23 \pm 0.04$ & $2.2 \pm 0.1$
      \\
      Vienna  & $2.20 \pm 0.04$ & $2.20 \pm 0.03$
     \\
     Walnut Creek  & $2.6 \pm 0.1$ & $2.4 \pm 0.1$
     \\
      Washington & $2.25 \pm 0.04$ & $2.24 \pm 0.03$
    \end{tabular}
    \caption{The market competition dimension $m_d$ for $19$ different cities from all over the world. The \textbf{$N(r)^{1/m_d}$ fit} column shows the results obtained by fitting how the maximum of the equilibrium average  price $\langle p^* \rangle$ scales with $N$, the number of nodes in the network. As a consistency check we report in \textbf{$N_{max}^{1/m_d}$ fit} column the values of the market dimension obtained simply from $\langle p^* \rangle = N_{max}^{1/m_d}$, where for each city $N_{max} = N(r=7)$ is the maximum order of the sampled graph, corresponding to a neighbourhood of range $r=7$ of the node with the highest closeness centrality.}
    \label{tab:cities_md}
\end{table} 
The results show a very good agreement between the two methods for all cities, despite the brute force method being obviously less refined. We recall that $m_d$ for a chain is $m_d^{chain} = 1$, while for a regular square lattice is equal $2$. We notice that for the all cities the fitted value of $m_d$ is greater than $2$. In particular for the cities in our dataset we found $2.1 \leq m_d \leq 3.14$.

\section{Hotelling centrality and average distance from the closest seller on a chain}

\begin{figure}[htp]
    \includegraphics[width=0.42\textwidth]{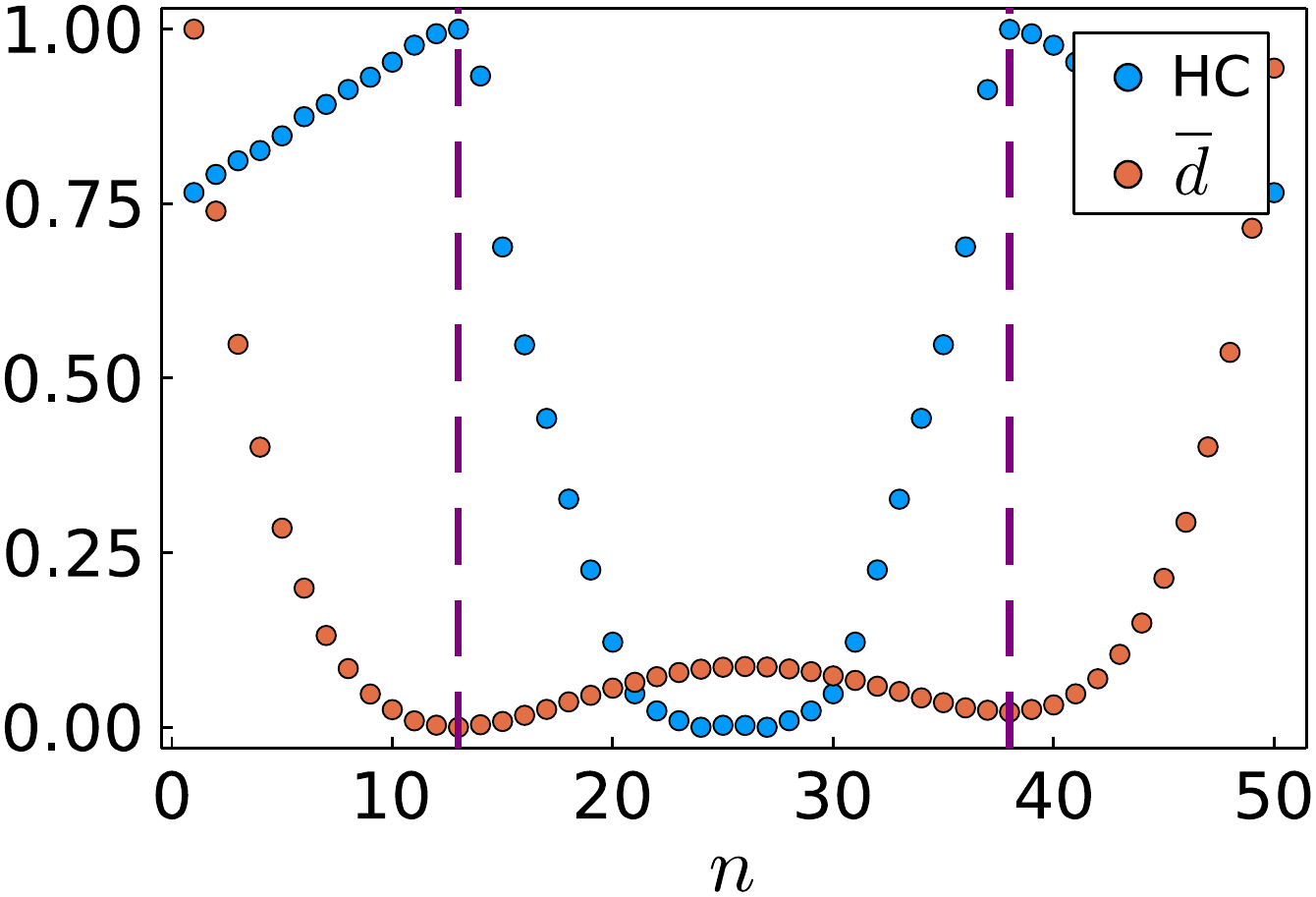}
  \caption{HC for BR price update compared to $\overline{d}$,  for the nodes of a chain of cardinality $N=50$. The two dashed lines indicate the two nodes $n = 13$ and $n = 38$ which at the same time maximize HC and minimize $\overline{d}$. }
\label{fig:HCvsavedistance}
\end{figure}

In this section we show that on a linear market represented as a chain graph, the nodes which maximise the Hotelling centrality (HC) under a BR dynamics are those which minimise $\overline{d}_n$, the average distance of the buyers from the closest seller given the position $n$ of one of the two sellers. For example we can focus on seller $\alpha$, and hence $\overline{d}_{n_{\alpha}}$ is the average transportation cost (measured in units of distance) of the buyers toward the closest seller, as a function of the node occupied by seller $\alpha$.
That is:
\begin{equation}
    \overline{d}_{n_{\alpha}} = \sum_{n_{\beta}} \sum_j \left[
    d_{n_{\alpha} n_j} \frac{\delta(min(d_{n_{\alpha} n_j}, d_{n_{\beta}}) - d_{n_{\alpha} n_j} )}{\sum_j \delta(min(d_{n_{\alpha} n_j}, d_{n_{\beta} n_j}) - d_{n_{\alpha} n_j} )} 
    +  d_{n_{\beta} n_j} \frac{\delta(min(d_{n_{\beta} n_j}, d_{n_{\alpha} n_j}) - d_{n_{\beta} n_j} )}{\sum_j \delta(min(d_{n_{\beta} n_j}, d_{n_{\alpha} n_j}) - d_{n_{\beta} n_j} )} \right]
\end{equation}
where $\delta(\cdot)$ is the Kronecker's delta.
Fig.~\ref{fig:HCvsavedistance} reports the values of HC for BR price update and $\overline{d}$ as a function of the node label $n$ for a chain of $N = 50$ nodes, where as usual the nodes are labeled using integer numbers in increasing order.
Both HC and $\overline{d}$ are normalized between $0$ and $1$.
We observe that the two nodes $n = 13$, $n = 38$ (i.e. corresponding respectively to $n/N \approx 0.25, 0.75$) maximize HC while at the same time minimize $\overline{d}$.
We recall that from the definition of HC (see main text) it follows that nodes with the highest HC are the best positions that a seller can occupy to maximize its earning, when the seller has no information on the position of the other seller. 
This means that on a chain market the positions which guarantee the average highest payoff to a seller are also the ones minimizing the transportation cost of the buyers. 
\end{document}